\documentclass[superscriptaddress,nofootinbib,amsmath,amssymb,
 aps,prd]{revtex4-2}

\usepackage{graphicx,subfigure,ulem}
\usepackage{dcolumn}
\usepackage{bm}
\usepackage{natbib} 
\usepackage[colorlinks=true,citecolor=blue,linkcolor=blue,urlcolor=blue]{hyperref}
\setcitestyle{square, numbers, comma, sort&compress} 
\usepackage{multirow}
\newcommand\mathplus{+}

\begin{document}
\title{Cold and dense perturbative QCD in a very strong magnetic background}

\author{Eduardo S. Fraga}
\email{fraga@if.ufrj.br}
\affiliation{Instituto de F\'isica, Universidade Federal do Rio de Janeiro, Caixa Postal 68528, 21941-972, Rio de Janeiro, RJ, Brazil}

\author{Let\'icia F. Palhares}%
 \email{leticia.palhares@uerj.br}
 \affiliation{Universidade do Estado do Rio de Janeiro, Instituto de F\'isica, Departamento de F\'isica Te\'orica, Rua S\~ao Francisco Xavier 524, 20550-013 Maracan\~a, Rio de Janeiro, Brasil}
 
\author{Tulio E. Restrepo}
 \email{tulioeduardo@pos.if.ufrj.br}

\affiliation{Instituto de F\'isica, Universidade Federal do Rio de Janeiro, Caixa Postal 68528, 21941-972, Rio de Janeiro, RJ, Brazil}

\begin{abstract}
We compute the pressure from first principles within perturbative QCD at finite baryon density and very high magnetic fields up to two-loops and with physical quark masses. The region of validity for our framework is given by $m_s \ll \mu_q \ll \sqrt{eB}$, where $m_s$ is the strange quark mass, $\mu_q$ is the quark chemical potential, $e$ is the fundamental electric charge, and $B$ is the magnetic field strength. We include the effects of the renormalization scale in the running coupling, $\alpha_s (\mu_q,\sqrt{eB})$, and running strange quark mass. We also discuss the simplifications that come about in the chiral limit. The effectively negligible contribution of the exchange diagram allows for building a simple analytic model for the equation of state for pure quark magnetars and computing their mass and radius at very large values of $B$. These results provide constraints on the
behavior of the maximum mass and associated radius from perturbative QCD. We also discuss the magnetic bag model for extreme magnetic fields.
\end{abstract}

\maketitle

\section{Introduction}

The phase diagram of strongly interacting matter matter under the influence of different external control parameters, such as temperature, different types of chemical potential and electromagnetic fields, is determined by in-medium quantum chromodynamics (QCD), its fundamental theory. The case of magnetic QCD, where one of the control parameters is an external magnetic field, is phenomenologically relevant in different scenarios, such as the astrophysics of compact stars \cite{Duncan:1992hi,Thompson:1993hn,Kouveliotou:1998ze}, in non-central, high-energy heavy ion collisions \cite{Kharzeev:2007jp,Skokov:2009qp,Voronyuk:2011jd,Bzdak:2011yy,Deng:2012pc,Inghirami:2016iru,Roy:2017yvg}, and in the early universe \cite{Vachaspati:1991nm,Enqvist:1993np,Grasso:2000wj}.

In particular, the thermodynamics of strong interactions in cold and dense matter under the influence of very strong magnetic fields is relevant to the description of magnetars \cite{Kaspi:2017fwg}. These objects correspond to a class of compact stars \cite{Schaffner-Bielich:2020psc} whose magnetic fields can be of the order of $10^{15}$ Gauss \cite{Duncan:1992hi,Thompson:1993hn,Kouveliotou:1998ze} at the surface, and possibly much higher in the core (up to $10^{20}$ Gauss \cite{Ferrer:2010wz}). The key ingredient to describe the microphysics and to compute the structural properties of magnetars is given by the equation of state for hadronic matter under such extreme conditions.
The NICER X-ray determination of mass and radius from millisecond pulsars PSR J0030+0451 \cite{Riley_2019,Miller:2019cac} and PSR J0740+6620 \cite{Riley:2021pdl,Miller:2021qha}, along with the LIGO-VIRGO detection of gravitational waves from binary neutron star mergers \cite{LIGOScientific:2018cki,LIGOScientific:2018hze}, constrained the equation of state that describes quark stars and neutron stars (see Ref. \cite{MUSES:2023hyz} for a review). 

In this paper we investigate the behavior of the pressure from first principles within perturbative QCD at finite density and very high magnetic fields up to two-loops for $N_f=3$ flavors with physical quark masses. We show that the exchange contribution increases with the magnetic field, but nevertheless corresponds to a correction of less than 3$\%$ at intermediate values of the quark chemical potential ($\mu_q \sim 300$ MeV) even for extremely large magnetic fields. Since we use perturbative QCD within the lowest-Landau level (LLL) approximation, the region of validity for our framework is given by $m_s \ll \mu_q \ll \sqrt{eB}$, where $m_s$ is the strange quark mass, $\mu_q$ the quark chemical potential, $e$ is the fundamental electric charge, and $B$ is the magnetic field strength.
In the case of symmetric quark matter, we consider values of the magnetic field $eB \sim 1- 9$ $\rm{GeV}^2$ ($B \sim 1.7-15.3 \times 10^{20}$ Gauss). These ranges ensure the applicability of the LLL approximation for relevant values of chemical potential ($\mu_q\leq 0.8$ GeV), and $9$ $\rm{GeV}^2$ corresponds to the highest value achieved by Lattice simulations for thermal QCD \cite{DElia:2021yvk}.
We also include the effects of the renormalization scale in the running coupling, $\alpha_s (\mu_q,\sqrt{eB})$, and running quark masses, since these effects have proven to be relevant for the resulting thermodynamics \cite{Freedman:1977gz,Farhi:1984qu,Fraga:2004gz,Kurkela:2009gj,Gorda:2021gha}. Finally, we discuss the simplifications that come about in the chiral limit, as was previously observed in the thermal magnetic QCD case \cite{Blaizot:2012sd,Fraga:2023cef}. 

Cold and dense magnetic QCD has a major difference with respect to its analogous at finite temperature and zero baryon density: it suffers from the Sign Problem \cite{Aarts:2015tyj}, so that it cannot be tackled by Monte Carlo simulations in the parameter sector that is relevant for compact star physics, which would correspond to large values of the baryon chemical potential. Therefore, one does not have lattice QCD as a benchmark to compare to. Besides magnetic pQCD, one can approach the equation of state in the limit of a large number of colors $N_c$ \cite{Fraga:2012ev}, via chiral perturbation theory \cite{Colucci:2013zoa}, using holographic models \cite{Preis:2010cq,Preis:2011sp} and, of course, within effective models. For a detailed discussion and list of references, see Refs. \cite{Fraga:2012rr,Kharzeev:2013jha,Andersen:2014xxa,Miransky:2015ava}. Here we ignore the effect of color superconductivity \cite{Alford:2007xm} in the presence of a strong magnetic field \cite{Ferrer:2005vd,Ferrer:2006vw,Ferrer:2007iw,Fukushima:2007fc,Noronha:2007wg}, which might be relevant to transport phenomena in the core of neutron stars and magnetars.

The equation of state for a system composed by up, down and strange quarks at zero temperature and nonzero baryon chemical potential, which we will refer to as cold and dense quark matter, was first obtained within perturbative QCD more than four decades ago by Freedman and McLerran \cite{Freedman:1976ub,Freedman:1977gz}, and also by Baluni \cite{Baluni:1977ms} and Toimela \cite{Toimela:1984xy}. Since then, it has been systematically improved \cite{Blaizot:2000fc,Fraga:2001id,Fraga:2004gz,Kurkela:2009gj,Fraga:2013qra,Kurkela:2014vha,Fraga:2015xha,Ghisoiu:2016swa,Annala:2017llu,Gorda:2018gpy,Annala:2019puf,Gorda:2021kme}. One relevant feature of perturbative QCD for cold and dense quark matter is that it seems to be much better behaved, compared to its thermal counterpart, in terms of the convergence of the series \cite{Braaten:2002wi}. Nevertheless, if we consider the exchange contribution, including the renormalization group running of $\alpha_s$ and $m_s$, it brings corrections $\sim 30\%$ for $\mu_q \sim 600$ MeV \cite{Fraga:2004gz}. Therefore, the sizable reduction in the exchange contribution in the presence of a strong magnetic background mentioned before has remarkable effects on the perturbative series, as discussed in Ref. \cite{Fraga:2023cef} for the thermal case.

The effectively negligible contribution of the exchange diagram allows for building a simple (analytic) description for the high-density sector of the equation of state for magnetic cold quark matter from perturbative QCD, provided that the magnetic background is strong enough to justify the lowest-Landau level description. We illustrate the utility of this formulation by computing some features of pure quark magnetars (strange magnetars) assuming extremely large magnetic fields. We also provide some estimates using an effective magnetic bag model that is partially justified in the chiral limit (also assuming a very high magnetic field) by the behavior of the magnetic exchange diagram. In this paper we restrict our analysis to pure quark magnetars since our point is just to illustrate the behavior of the equation of state derived from magnetic perturbative QCD and provide boundaries obtained from the fundamental theory of strong interactions.

To compute hybrid magnetars or magnetized proto neutron stars, however, one has to include the low-density sector and perform a matching of the equations of state (see Refs. \cite{Rabhi:2009ih,Dexheimer:2011pz,Dexheimer:2012mk,Franzon:2015sya,Rather:2022bmm}). Moreover, one should also include crust effects \cite{Fang:2016kcm,Fang:2017zsb}. To consider magnetic fields typical of magnetars, one should also go beyond the lowest-Landau level approximation in the pQCD sector of the equation of state, which is very challenging. This more realistic description of the star composition is however out of the scope of the current presentation. Here we focus on computing the equation of state from the fundamental theory of strong interactions within a specific region of validity where our approximations are under control. 

The properties of strange quark magnetars have so far been mostly studied within the framework of effective theories, such as the Nambu Jonna-Lasinio model \cite{Menezes:2008qt,Menezes:2009uc,Menezes:2014aka,Peterson:2023bmr}, the density-dependent quark model \cite{Anand_2000,Wei_2017}, the quasiparticle model \cite{Wen:2012jw,Chu:2018dch,Zhang:2021qhl}, the chiral SU(3) quark mean field model \cite{Kumari:2022hyx} and the bag model \cite{Dexheimer:2012mk,PeresMenezes:2015ukv,Kayanikhoo:2019ugo,Ferreira:2022fjo}.

This work is organized as follows. In Section \ref{sec:pressure} we present the perturbative setup and a few details on the calculation of the pressure to two-loops, including the running of the coupling and strange quark mass. In Section \ref{sec:results} we discuss our results for the pressure. In Section \ref{sec:bag}, we present some estimates from an effective magnetic bag model, which can be justified in the chiral limit. In Section \ref{sec:pQCD-model} we propose an effective pQCD model for the equation of state and compute some features of quark magnetars. Section \ref{sec:outlook} contains our summary and outlook.

\section{Pressure in cold and dense magnetic perturbative QCD}
\label{sec:pressure}

In this section we compute the pressure in perturbative QCD to two-loops. We assume that the system is embedded in a uniform, {\it very} large magnetic field ${\bf B}=B \hat{\bf z}$, where the field strength $B$ is much larger than the chemical potential and all masses.


Let us start with the one-loop (free), contribution to the pressure of in-medium QCD in the presence of high magnetic fields. Since we will consider the case with zero temperature, there is only the contribution coming from the quark sector, which is given by the following renormalized expression (subtracting the pure vacuum term) in the lowest Landau level (LLL)\cite{Fraga:2012rr,Fraga:2012fs,Blaizot:2012sd,Kharzeev:2013jha,Andersen:2014xxa,Miransky:2015ava,Fraga:2023cef}:
%
\begin{widetext}
\begin{align}
\begin{split}
 \frac{P_{\rm free}^{\rm LLL}}{N_c}=&
 -\sum_f\frac{(q_fB)^2}{2\pi^2}\left[x_f\ln\sqrt{x_f}\right]+T\sum_{f}\frac{q_fB}{2\pi}\int \frac{dp_z}{2\pi}\bigg \{\ln\left(1+e^{-\beta\left[E(p_z)-\mu_f\right]}\right)+\ln\left(1+e^{-\beta\left[E(p_z)+\mu_f\right]}\right)\bigg \} \, ,
 \end{split} \label{Pfree}
\end{align}
\end{widetext}
%
where $N_c=3$, $E=\sqrt{p_z^2+m_f^2}$, $x_f\equiv m_f^2/2q_f B$ and $q_f, m_f, \mu_f$ are respectively the absolute value of the electric charge, the mass and the chemical potential of the $f$-quark. In the zero-temperature limit $\beta\to \infty$, Eq. (\ref{Pfree}) has a simpler form in terms of the Heaviside $\Theta$ function (for details, see Ref. \cite{Palhares:2012fv})
\begin{widetext}
\begin{align}
 \frac{P_{\rm free}^{\rm LLL}}{N_c}=&
 -\sum_f\frac{(q_fB)^2}{2\pi^2}\left[x_f\ln\sqrt{x_f}\right]+\sum_{f}\frac{q_fB}{2\pi}\int \frac{dp_z}{2\pi}\left(\mu_f-E\right)\Theta(\mu_f-E)\\
 =& -\sum_f\frac{(q_fB)^2}{2\pi^2}\left[x_f\ln\sqrt{x_f}\right]+\sum_{f}\frac{q_fB}{2\pi}\int_0^{P_F} \frac{dp_z}{2\pi}\left(\mu_f-E\right)\\
=& -\sum_f\frac{(q_fB)^2}{2\pi^2}\left[x_f\ln\sqrt{x_f}\right]+\sum_{f}\frac{(q_fB)}{4\pi^2}\left[ \mu_f P_F - m_f^2 \log\left( \frac{\mu_f+P_F}{m_f} \right) \right] 
 \, ,
  \label{Pfree_T0}
\end{align}
\end{widetext}
where $P_F=\sqrt{\mu_f^2-m_f^2}$ is the Fermi momentum. Notice that, in principle, the first (magnetic vacuum) term contributes for all values of $\mu_f$, whereas the second term appears only for $\mu_f > m_f$, as usual.
It is easy to check from the Heaviside function, that the next Landau level ($n=1$), only contributes if $B<\frac{\mu_f^2-m_f^2}{2q_f}$.

In the chiral limit, we have simply
\begin{equation}
\left[\frac{P_{\rm free}^{\rm LLL}}{N_c}\right]_{\rm chiral}= \frac{q_f B}{4\pi^2}\mu_f^2 \,.   
\label{free-chiral}
\end{equation}
%

\begin{figure*}[!ht]

 \begin{subfigure}
 \centering
 \includegraphics[width=0.45\textwidth]{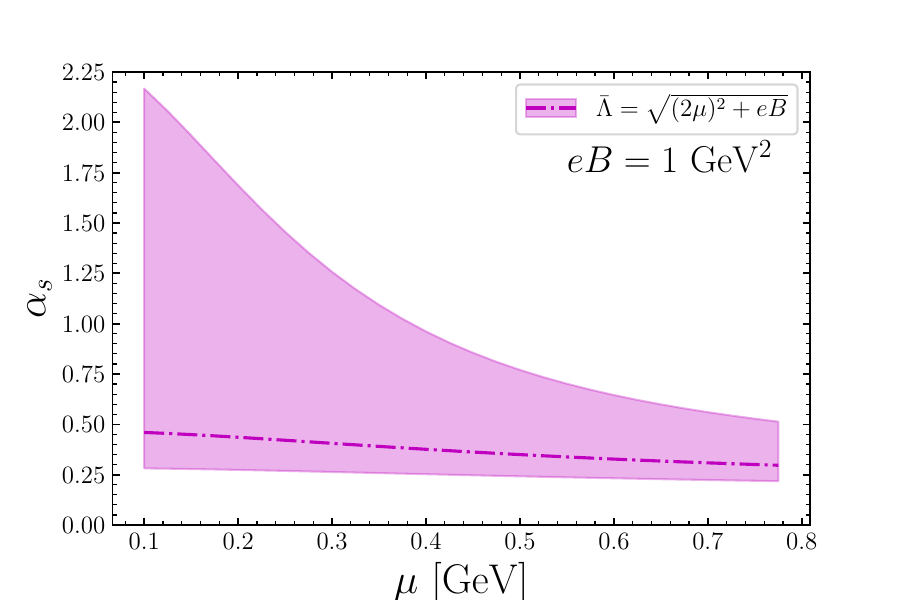} 
 \end{subfigure}
 \begin{subfigure}
 \centering
 \includegraphics[width=0.45\textwidth]{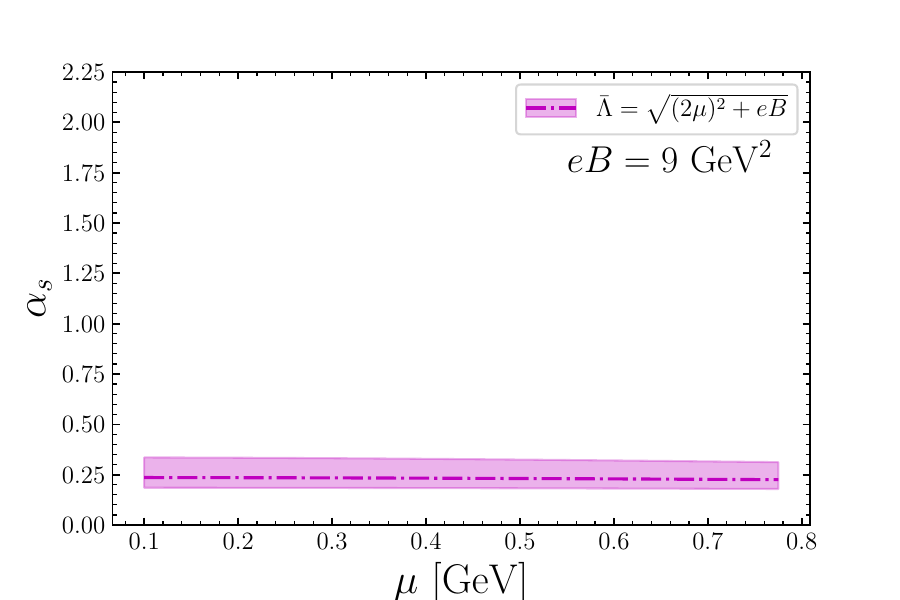} 
 \end{subfigure}
 
\begin{subfigure}
 \centering
 \includegraphics[width=0.45\textwidth]{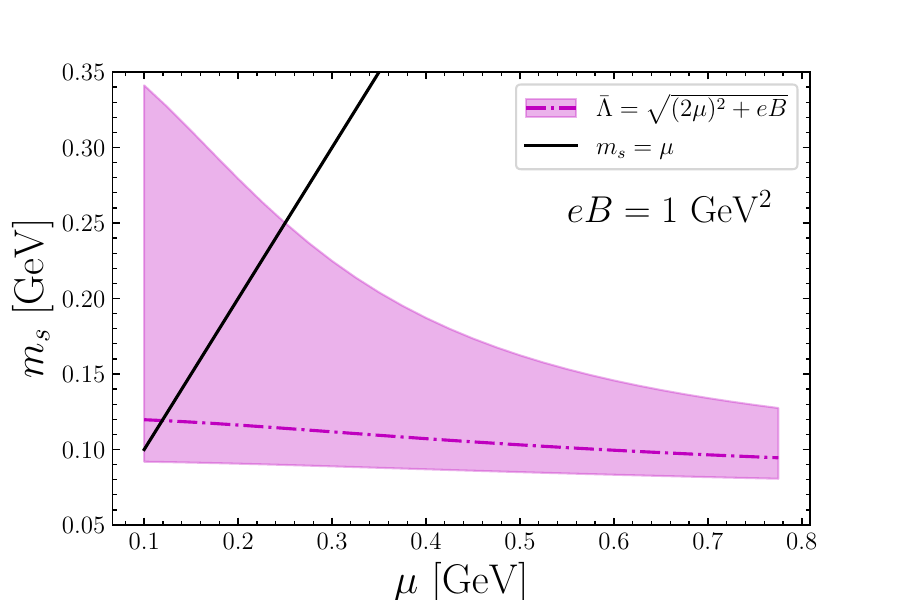} 
 \end{subfigure}
 \begin{subfigure}
 \centering
 \includegraphics[width=0.45\textwidth]{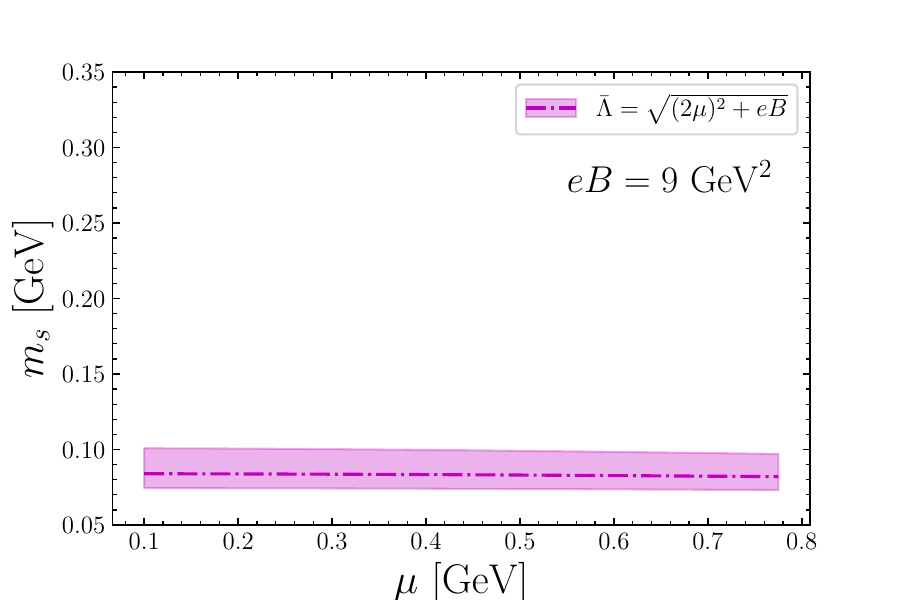} 
 \end{subfigure}

\caption{Running coupling (top) and strange quark mass (bottom) as functions of the chemical potential for $eB=1$ $\rm{GeV}^2$ (left) and $eB=9$ $\rm{GeV}^2$ (right). The bands correspond to changes in the central scale by a factor of $2$. For these plots $\mu=\mu_u=\mu_d=\mu_s$, assuming symmetric matter.}
\label{fig:ms}
\end{figure*}

The two-loop pressure at finite density and temperature  was first computed in Ref. \cite{Blaizot:2012sd} in the LLL approximation. It corresponds to the exchange diagram and has the following form (for one flavor):
\begin{align}
\begin{split}
 \frac{P_{\rm exch}^{\rm LLL}}{N_c}=&-\frac{1}{2}g^2 \left(\frac{N_c^2-1}{2N_c}\right) m_f^2\left(\frac{q_fB}{2\pi}\right)\int \frac{dm_k}{2\pi}m_ke^{-\frac{m_k^2}{2 q_f B}}\int \frac{dp_zdq_zdk_z}{(2\pi)^3}(2\pi)\delta(k_z-p_z+q_z)\frac{1}{\omega E_p E_q}\Bigg\{\frac{\omega\Sigma_+}{E_-^2-\omega^2}\\
 &+\frac{\omega\Sigma_-}{E_+^2-\omega^2}+2\left[\frac{E_+}{E_+^2-\omega^2}-\frac{E_-}{E_-^2-\omega^2}\right]n_B(\omega)N_F(1)-\left[\frac{2\left(E_{\bf{q}}+\omega\right)}{\left(E_--\omega\right)\left(E_++\omega\right)}\right]N_F(1)\\
 &-2\frac{E_+}{E_+^2-\omega^2}n_B(\omega)-\frac{1}{E_{+}+\omega}\Bigg\},
\end{split}\label{P_exch_T_finite-0}
\end{align}
where
\begin{align}
\begin{split}
     E_\pm\equiv& E_{\bf{p}}\pm E_{\bf{q}}=\sqrt{{\bf{p}}^2+m_f^2}\pm\sqrt{{\bf{q}}^2+m_f^2},\\
     N_F(1)=&n_F(E_{\bf{p}}+\mu_f)+n_F(E_{\bf{p}}-\mu_f),\\
     \Sigma_\pm\equiv & n_F(E_{\bf{p}}+\mu_f)n_F(E_{\bf{q}}\pm\mu_f)+n_F(E_{\bf{p}}-\mu_f)n_F(E_{\bf{q}}\mp\mu_f).
\end{split}
\end{align}
Here, $n_B$ and $n_F$ are the Bose-Einstein and Fermi-Dirac distributions, respectively, and $\omega=\sqrt{k_z^2+m_k^2}$, where $m_k^2$ accounts for the gluon momentum components transverse to the magnetic field. Again, in the limit $T\to 0$, we have $n_F(E_{\bf{p}}+\mu_f)\to 0$, $n_F(E_{\bf{p}}-\mu_f)\to \Theta(\mu_f-E_{\bf{p}})$ and $n_B(\omega)\to 0$, so that $N_F(1)\to \Theta(\mu_f-E_{\bf{p}}) $, $\Sigma_-\to 0 $ and $\Sigma_+\to \Theta(\mu_f-E_{\bf{p}})\Theta(\mu_f-E_{\bf{q}})$. Then Eq. (\ref{P_exch_T_finite-0}) acquires the form 
\begin{align}
\begin{split}
 \frac{P_{\rm exch}^{\rm LLL}}{N_c}=&-\frac{1}{2}g^2 \left(\frac{N_c^2-1}{2N_c}\right) m_f^2\left(\frac{q_fB}{2\pi}\right)\int \frac{dm_k}{2\pi}m_ke^{-\frac{m_k^2}{2 q_f B}}\int \frac{dp_zdq_zdk_z}{(2\pi)^3}(2\pi)\delta(k_z-p_z+q_z)\\
 &\times\frac{1}{\omega E_p E_q}\Bigg\{\frac{\omega}{E_-^2-\omega^2}\Theta(\mu_f-E_{\bf{p}})\Theta(\mu_f-E_{\bf{q}}) -\left[\frac{2\left(E_{\bf{q}}+\omega\right)}{\left(E_--\omega\right)\left(E_++\omega\right)}\right]\Theta(\mu_f-E_{\bf{p}})-\frac{1}{E_{+}+\omega}\Bigg\} \, .
\end{split}\label{P_exch_T_finite}
\end{align}

The pressure to two-loops for three flavors ($N_f=3$) with physical quark masses depends not only on the chemical potential and magnetic field, but also on the renormalization subtraction point $\bar{\Lambda}$, an additional mass scale generated by the truncation in the perturbative expansion. 
This comes about via the scale dependence of both the strong coupling $\alpha_s(\bar{\Lambda})$ and strange quark mass $m_s(\bar{\Lambda})$.

The running of both $\alpha_s$ and $m_s$ are known to four-loop order in the $\overline{\rm MS}$ scheme \cite{Vermaseren:1997fq}. Since we have determined the pressure only to first order in $\alpha_s=g^2/4\pi$, we use for the coupling \cite{Fraga:2004gz}
     \begin{equation}
     \alpha_{s}(\bar{\Lambda})=\frac{4\pi}{\beta_{0}L}\left(
     1-\frac{2\beta_{1}}{\beta^{2}_{0}}\frac{\ln{L}}{L}\right) \,,
     \label{eq:alphas}
     \end{equation}
where $\beta_{0}=11-2N_{f}/3$, $\beta_{1}=51-19N_{f}/3$, $L=2\ln\left(\bar{\Lambda}/\Lambda_{\rm \overline{MS}}\right)$. Since $\alpha_{s}$ depends on $N_{f}$, fixing the massive quark at some energy scale also depends on the number of flavors.
For the strange quark mass, we have
	\begin{eqnarray}
	\begin{aligned}
	m_{s}(\bar{\Lambda})=\hat{m}_{s}\left(\frac{\alpha_{s}}{\pi}\right)^{4/9}
	\left[1+0.895062\left(\frac{\alpha_{s}}{\pi}\right)  
 \right] \;,
	\end{aligned}
	\label{eq:smass}
	\end{eqnarray}
with $\hat{m}_{s}$ being the renormalization group invariant strange quark mass, i.e. $\bar{\Lambda}$ independent. Since Eq. (\ref{eq:alphas}) for $\alpha_{s}$ tells us that different values of $N_{f}$ give different values of $\Lambda_{\overline{\rm MS}}$, by choosing  $\alpha_{s}(\bar{\Lambda}=1.5~{\rm GeV},~N_{f}=3)=0.336^{+0.012}_{-0.008}$ \cite{Bazavov:2014soa}, we obtain $\Lambda^{2+1}_{\overline{\rm MS}}=343^{+18}_{-12}~$MeV. Fixing the strange quark mass at $m_{s}(2~{\rm GeV}, N_{f}=3)=92.4(1.5)~$MeV \cite{Chakraborty:2014aca} gives $\hat{m}^{2+1}_{s}~{\approx}~248.7~$MeV when using $\alpha^{2+1}_{s}$ in Eq. (\ref{eq:smass}).

\begin{figure*}[!ht]
\begin{subfigure}
 \centering
 \includegraphics[width=0.45\textwidth]{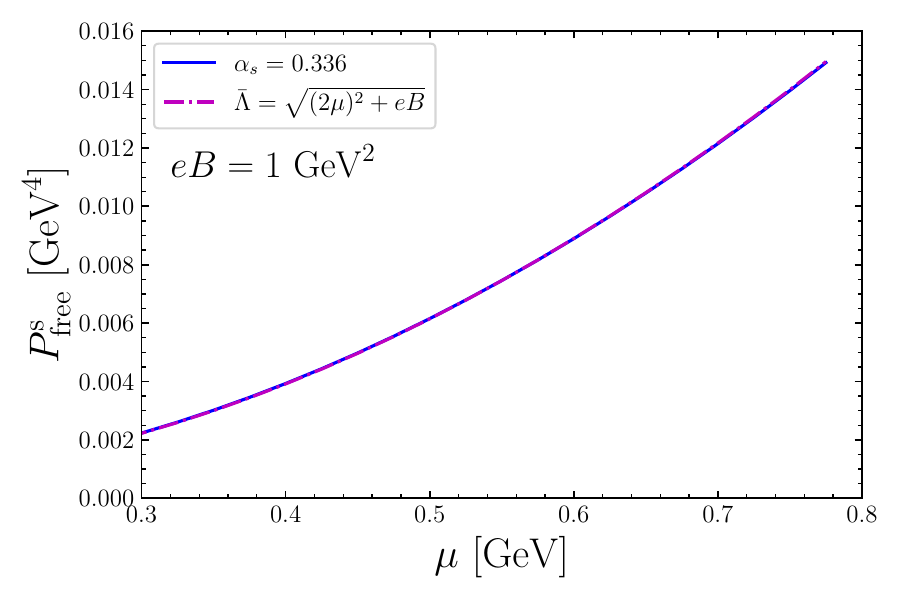} 
 \end{subfigure}
 \begin{subfigure}
  \centering
 \includegraphics[width=0.45\textwidth]{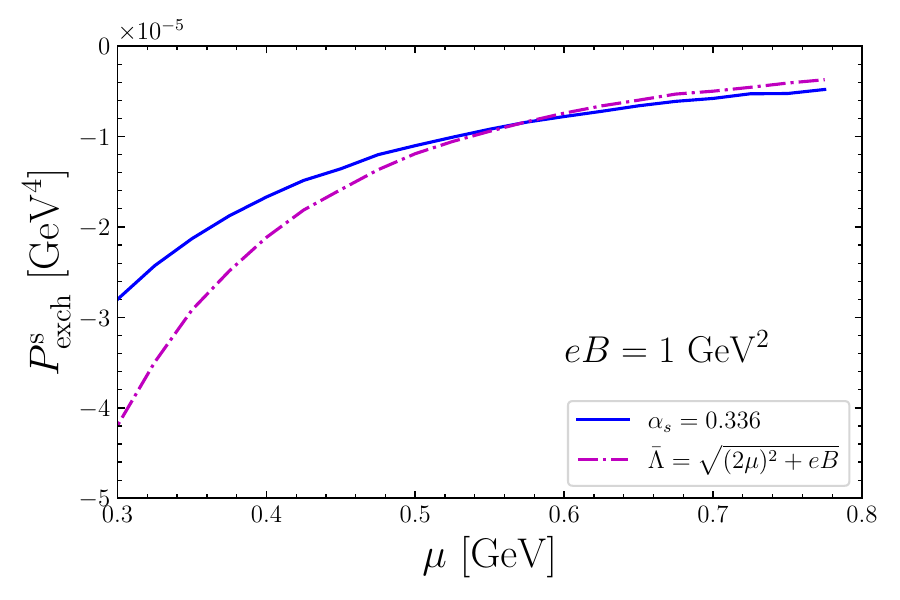}
 \end{subfigure}
\begin{subfigure}
  \centering
 \includegraphics[width=0.45\textwidth]{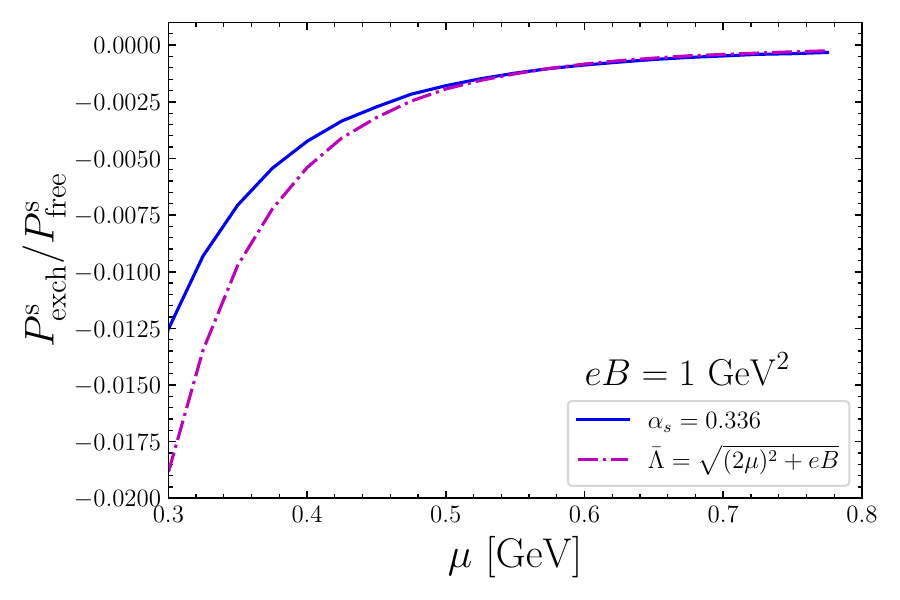}
 \end{subfigure}
 \begin{subfigure}
 \centering
 \includegraphics[width=0.45\textwidth]{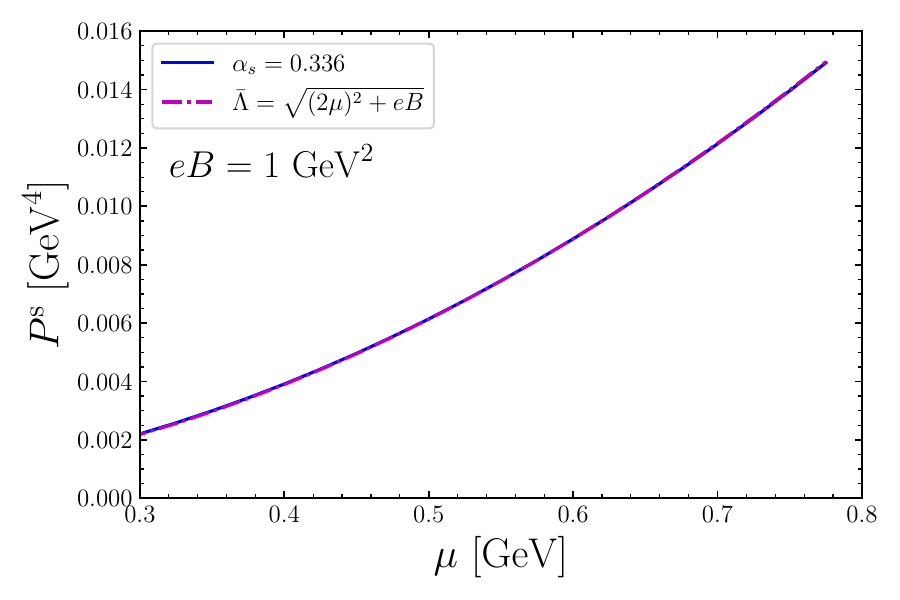} 
 \end{subfigure}
\caption{$P^s_{\rm free}$ (top-left), $P^s_{\rm exch}$ (top-right), $P^s_{\rm exch}/P^s_{\rm free}$ (bottom-left), and $P^s$ (bottom-right) as functions of the chemical potential for $eB=1$ $\rm{GeV}^2$.}
\label{fig:Ps_B1}
\end{figure*}

\begin{figure*}[!ht]
\begin{subfigure}
 \centering
 \includegraphics[width=0.45\textwidth]{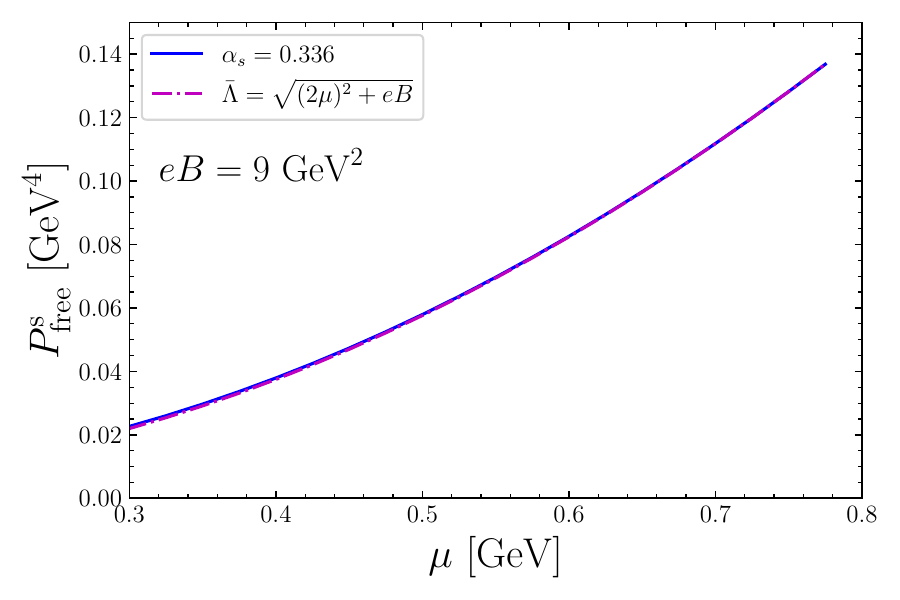} 
 \end{subfigure}
 \begin{subfigure}
  \centering
 \includegraphics[width=0.45\textwidth]{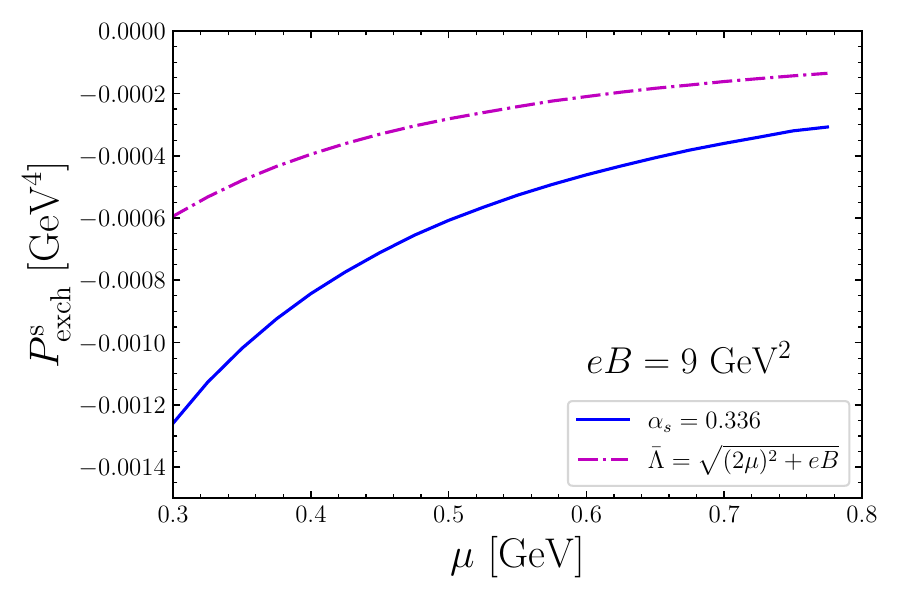}
 \end{subfigure}
\begin{subfigure}
  \centering
 \includegraphics[width=0.45\textwidth]{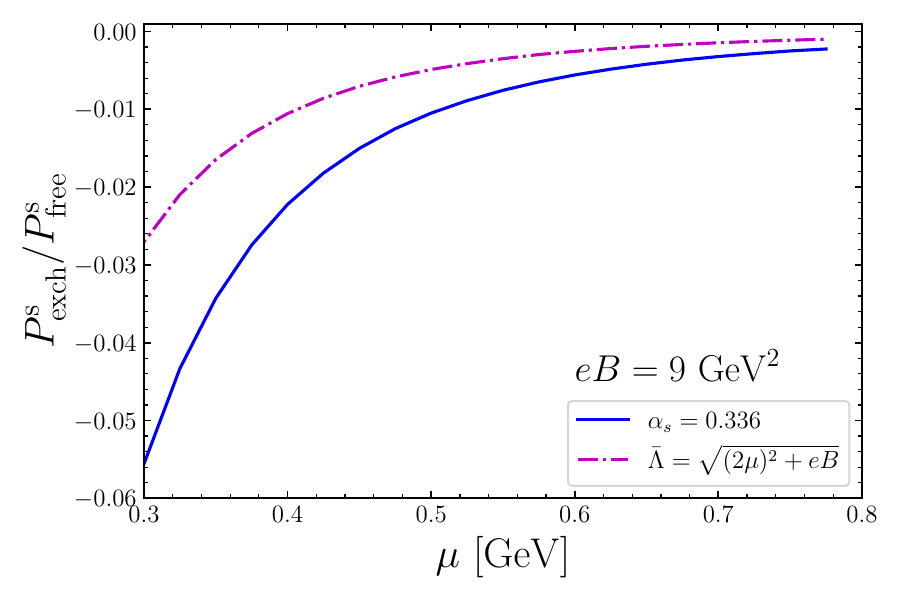}
 \end{subfigure}
 \begin{subfigure}
 \centering
 \includegraphics[width=0.45\textwidth]{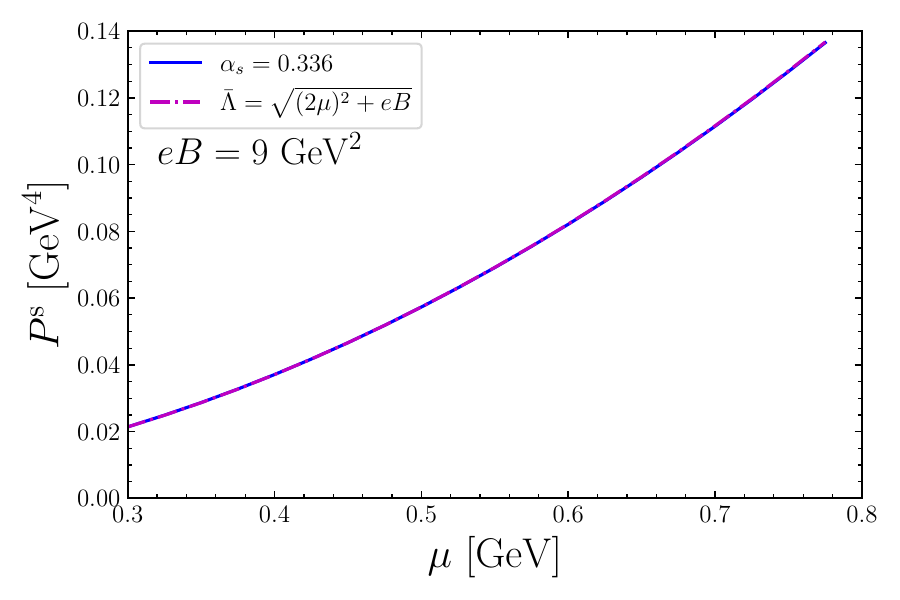} 
 \end{subfigure}
\caption{$P^s_{\rm free}$ (top-left), $P^s_{\rm exch}$ (top-right), $P^s_{\rm exch}/P^s_{\rm free}$ (bottom-left), and $P^s$ (bottom-right) as functions of the chemical potential for $eB=9$ $\rm{GeV}^2$.}
\label{fig:Ps_B9}
\end{figure*}

As usual, there is arbitrariness in the way one should connect the renormalization scale $\bar{\Lambda}$ to a physical mass scale of the system under consideration \cite{Kapusta:2006pm}. In cold and dense QCD where, besides quark masses, the only scale is given by the quark chemical potential, and $\mu_f\gg m_f$, the usual choice is $2\mu_f$ with a band around it, i.e. $\mu_f < \bar{\Lambda} < 4\mu_f$. In the present case, where the magnetic field also provides a relevant mass scale given by $\sqrt{eB}$, the choice becomes more ambiguous. Following the detailed analysis of different ways of implementing the scale dependence performed in Ref. \cite{Fraga:2023cef}, we show results only for the most physical case that emerged in that study: $\bar{\Lambda}=\sqrt{(2\mu_f)^2+eB}$, which corresponds to a natural extension of what is done in finite-temperature field theory at nonzero density \cite{Kapusta:2006pm}.

In Figure \ref{fig:ms} we show results for the renormalization group running of the strong coupling, $\alpha_s$, and the strange quark mass, $m_s$, as functions of the chemical potential $\mu=\mu_u=\mu_d=\mu_s$ (assuming symmetric matter) for $eB=1$ GeV$^2$ and $eB=9$ GeV$^2$, including bands that encode the renormalization-scale dependence in the standard range between half the central scale $\bar{\Lambda}=\sqrt{(2\mu_f)^2+eB}$ and two times it. To delimit the region of validity, we also show the line $m_s=\mu$ in the plots of the running of the strange quark mass. It is clear from the figures that, for large values of the magnetic field, the coupling remains small in all the relevant parameter region and the bands become narrower with increasing $B$. Notice also that both $\alpha_s$ and $m_s$ become very flat functions of $\mu$ for very large magnetic fields. In fact, at lowest order, one has $(\partial m_s/\partial\mu_s)=O(\alpha_s^{13/9}\mu_B/eB) \ll 1$ for large $B$.

\section{Results for the pressure}
\label{sec:results}

In this section, we present results for the pressure of symmetric quark matter as a function of the chemical potential $\mu=\mu_u=\mu_d=\mu_s$ for $eB=1$ GeV$^2$ and $eB=9$ GeV$^2$, the latter being the highest value of magnetic field attained in currently available lattice simulations in the case of thermal QCD \cite{DElia:2021yvk}. We also show results for the pressure as a function of the the magnetic field for $\mu=0.6$ GeV. Panels with the free pressure, $P^s_{\rm free}$, the exchange diagram contribution, $P^s_{\rm exch}$, the ratio $P^s_{\rm exch}/P^s_{\rm free}$, and the full strange pressure, $P^s$, are displayed. We concentrate on the contribution from the strange quark because mass effects are more relevant in this case. The ratio $P^s_{\rm exch}/P^s_{\rm free}$ provides a certain measure of the reliability of perturbation theory, since it seems to be more well behaved than the case in the absence of a large magnetic field \cite{Blaizot:2012sd}. Besides the case that includes the renormalization group running of $\alpha_s$ and $m_s$, we plot the case with no running for comparison.

For simplicity, we fully neglect anisotropy effects \cite{Huang2010,Bali:2013esa}. Results shown in this paper correspond to the longitudinal pressure in an anisotropic description \cite{Bali:2014kia}. One can show that significant anisotropies for quark magnetars appear for fields above $5\times 10^{18}$ Gauss \cite{Huang2010}. Nevertheless, we focus on the analysis of QCD interactions in the (longitudinal part of the) equation of state in this magnetic, cold and dense environment and recall that we do not aim at providing a realistic description of magnetars here.

In Figures \ref{fig:Ps_B1} and \ref{fig:Ps_B9} we plot the pressure as a function of the strange quark chemical potential for the two different values of external magnetic field. One can see that the effects from the running of the coupling and the mass are significant, reaching $\sim 50\%$ for smaller values of $\mu$ in the exchange contribution, when compared to the case with no running. On the other hand, for high magnetic fields it is clear that the two-loop contribution is comparatively negligible, representing a correction of a few percent. Even for much smaller fields, $eB \sim 0.0059~$GeV$^2$ ($B = 10^{18}~$Gauss), typical of magnetar cores, the obtained exchange contribution remains negligible. Still, one expects the LLL approximation to break down for such ``low'' field intensities. On the other hand, as will become clear in what follows, running effects always play a relevant role, since they also affect the leading term of the pressure via the running of the mass.

\begin{figure*}[!ht]
\begin{subfigure}
 \centering
 \includegraphics[width=0.45\textwidth]{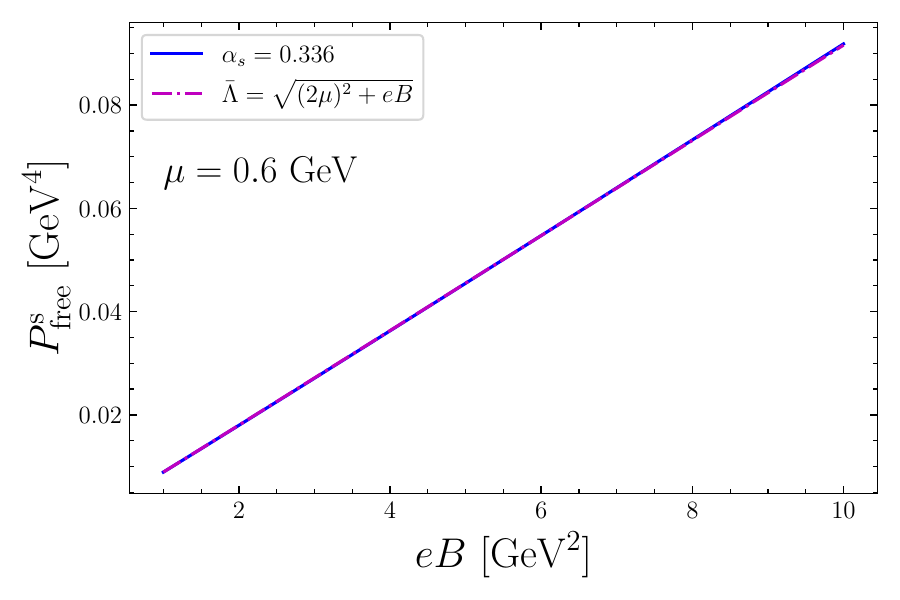} 
 \end{subfigure}
 \begin{subfigure}
  \centering
 \includegraphics[width=0.45\textwidth]{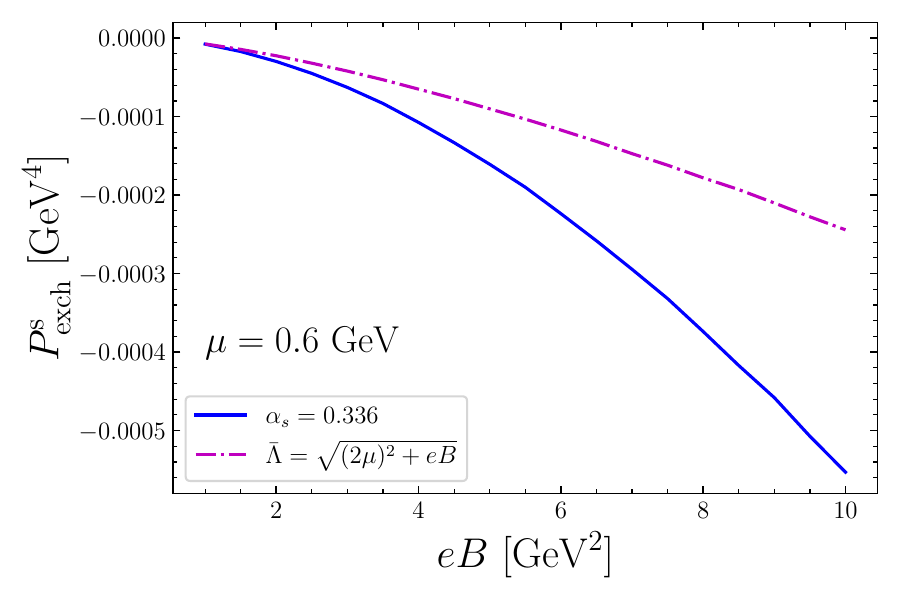}
 \end{subfigure}
\begin{subfigure}
  \centering
 \includegraphics[width=0.45\textwidth]{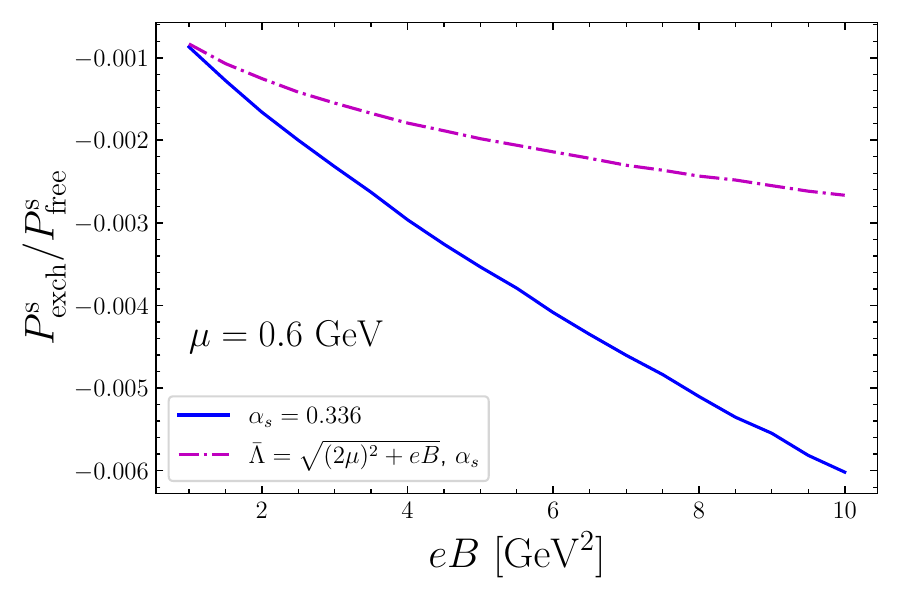}
 \end{subfigure}
 \begin{subfigure}
 \centering
 \includegraphics[width=0.45\textwidth]{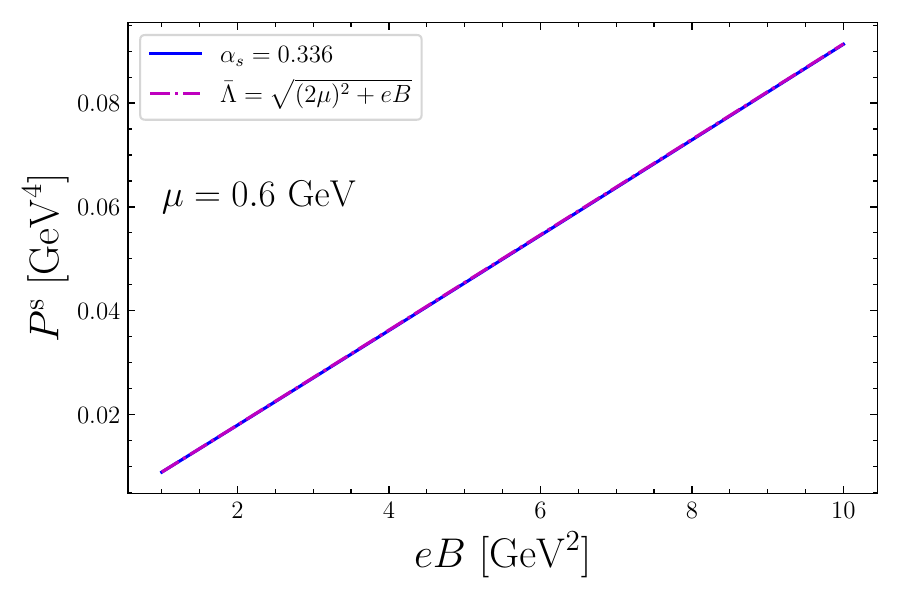} 
 \end{subfigure}
\caption{$P^s_{\rm free}$ (top-left), $P^s_{\rm exch}$ (top-right), $P^s_{\rm exch}/P^s_{\rm free}$ (bottom-left), and $P^s$ (bottom-right) as functions of the magnetic field for $\mu=0.6$ $\rm{GeV}$.}
\label{fig:Ps_mu_600}
\end{figure*}

In Figure \ref{fig:Ps_mu_600} we show the behavior of the pressure as a function of the magnetic field for $\mu_s=600~$ MeV. Although the exchange contribution increases with the magnetic field, it remains comparatively very small even for values of the field way beyond what is expected to be found in astrophysical environments.

\begin{figure*}[!ht]
\begin{subfigure}
 \centering
 \includegraphics[width=0.45\textwidth]{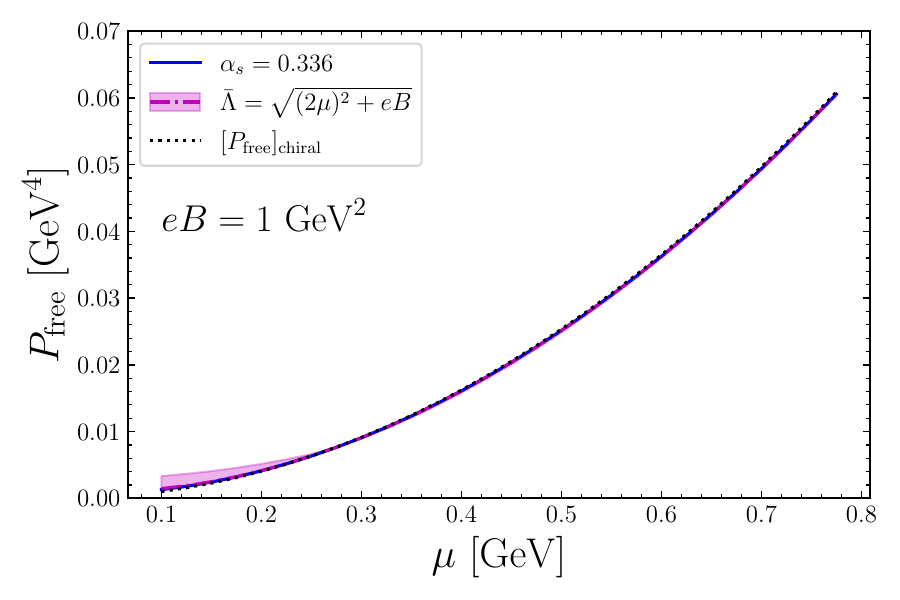} 
 \end{subfigure}
 \begin{subfigure}
 \centering
 \includegraphics[width=0.45\textwidth]{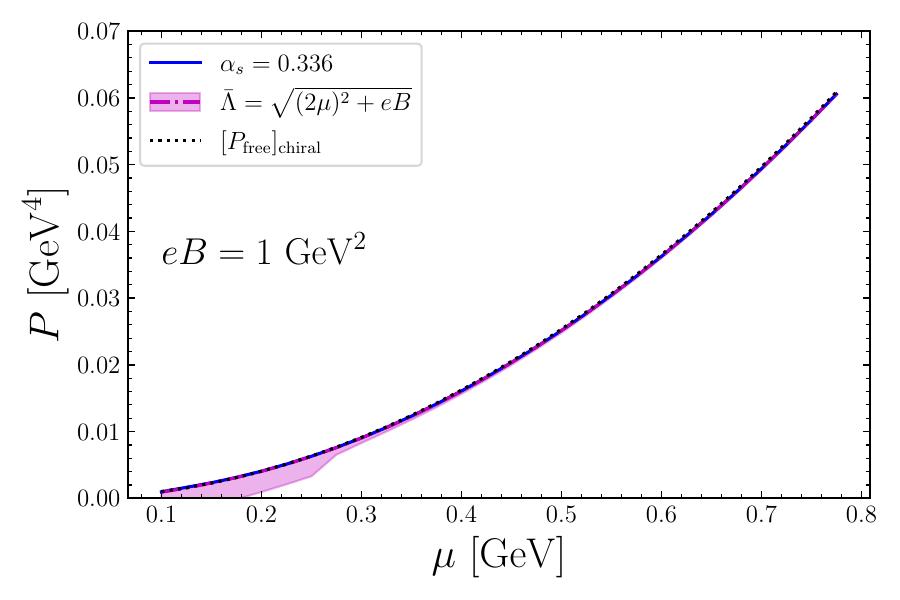} 
 \end{subfigure}
\caption{$P_{\rm free}$ (left) and $P$ (right) as functions of the chemical potential for $eB=1$ $\rm{GeV}^2$.  The bands correspond to changes in the central scale by a factor of $2$. For comparison, we also display the pressure in the chiral limit, given by Eq. (\ref{free-chiral})}
\label{fig:Ptot_B1}
\end{figure*}

Finally, we show in Figure \ref{fig:Ptot_B1} the free and the full pressures, both including renormalization group running bands between half and twice $\overline{\Lambda}$, for $eB=1$ GeV$^2$. For comparison, we also display the pressure in the chiral limit, given by Eq. (\ref{free-chiral}). One can see that, for large magnetic fields, all cases are essentially indistinguishable unless one goes to values of the chemical potential below $\mu_s=300~$ MeV. This fact suggests that one could build a pQCD-based simple analytic model for the equation of state of cold and dense quark matter under very large magnetic fields that could be used to describe quark magnetars. Before that, however, let us discuss some estimates from the magnetic bag model that are better justified after the results we have just discussed.

\section{Magnetic bag model and estimates for quark magnetars}
\label{sec:bag}

As discussed previously, the chiral limit for perturbative QCD at very high magnetic fields is extremely simple. The exchange contribution vanishes identically for $m_f=0$, as was already discussed in Refs. \cite{Blaizot:2012sd,Fraga:2023cef}, and the free contribution is given by Eq. (\ref{free-chiral}). We can use this fact to justify building an effective magnetic bag model by including only the purely magnetic contribution to the pressure and energy density, besides the usual bag constant. Magnetic versions of the bag model have been considered previously in \cite{Chakrabarty:1996te,Fraga:2012fs,Dexheimer:2012mk,PeresMenezes:2015ukv,Ferreira:2022fjo}. Here, however, we consider this model for very large magnetic fields, which brings the quadratic behavior of the pressure with the chemical potential shown in Eq. (\ref{free-chiral}). This has important consequences, as we will see, since the equation of state that results is much stiffer than in the usual MIT bag model, where $P_f^{\rm MIT}\sim \mu_f^4$.

Following the usual procedure for building a bag model \cite{Witten:1984rs,Alcock:1986hz,Weissenborn:2011qu,Schaffner-Bielich:2020psc,Sotani:2014rva,Sotani:2017qjo}, it is straightforward to derive the following equation of state from the thermodynamic potential:
\begin{equation}
P_{\rm bag}=\epsilon - 2\cal{B}_{\rm eff} \,, 
\label{eos-zel}
\end{equation}
where we have an effective bag constant ${\cal B}_{\rm eff}$ in the presence of the magnetic field $B$ given by 
\begin{equation}
{\cal B}_{\rm eff} \equiv {\cal B} + \frac{B^2}{2}\,,   
\end{equation}
${\cal B}$ being the bag constant. Here, we have essentially used Eq. (\ref{free-chiral}) and the first law of thermodynamics. This should be contrasted to the usual MIT bag model that yields the following equation of state:
\begin{equation}
P_{\rm MIT}=\frac{1}{3}(\epsilon - 4 {\cal B}) \,.  
\end{equation}

The new form of the equation of state, Eq. (\ref{eos-zel}), obtained in the limit of extremely high magnetic fields, is also a self-bound equation of state, but a much stiffer one. It corresponds to the Zel'dovich equation of state \cite{Zeldovich:1961sbr} and represents the limiting causal situation, with speed of sound $c_s=1$ \cite{Schaffner-Bielich:2020psc}. This comes about because, in the presence of an extremely strong magnetic field, one has $P_{\rm bag}\sim eB \mu_f^2$ instead of $P_{\rm MIT}\sim \mu_f^4$, as discussed above. 

We can use a scaling, analogous to the one originally proposed by Witten \cite{Witten:1984rs} in the case of the original MIT bag model, to write the maximum mass, radius and central energy density of a strange quark magnetar obtained from the Tolman-Oppenheimer-Volkov (TOV) equations \cite{Oppenheimer:1939ne}. Using the Zel'dovich equation of state above as an input, one obtains \cite{Schaffner-Bielich:2020psc}
\begin{equation}
M^{\rm max}=4.23M_\odot \left( \frac{{\cal \epsilon_{\rm sat}}}{{2\cal B}_{\rm eff}} \right)^{1/2}    \, ,
\end{equation}
\begin{equation}
R^{\rm max}=17.6 ~{\rm km} \left( \frac{{\cal \epsilon_{\rm sat}}}{{2\cal B}_{\rm eff}} \right)^{1/2}     \, ,
\end{equation}
\begin{equation}
\epsilon_c^{\rm max}= 3.03 ~(2 {\cal B}_{\rm eff})  \, ,   
\end{equation}
where $M_\odot$ is the solar mass and $\epsilon_{\rm sat} = 140~$MeV/fm$^3$ is the energy density at nuclear matter saturation. These equations only make sense for magnetic field strengths such that $\sqrt{eB} \gg \mu_f$. Here we adopt the value ${\cal B}^{1/4}=145~$MeV (${\cal B}=57~$MeV/fm$^3$) for the bag constant. 

A magnetic field of $B=10^{18}~$Gauss, for instance, typical of magnetar cores, corresponds to $\sqrt{eB}\sim 77~$MeV, which is ``too low''. For $B=2 \times 10^{19}~$Gauss, one has $\sqrt{eB}\sim 342~$MeV, which is of the same order as $\mu_f$ and yields an energy density of $\frac{B^2}{2}=900~$MeV/fm$^3$, so that ${\cal B}_{\rm eff}=957~$MeV/fm$^3$.
Then, we can estimate the values of the maximum mass, radius and central energy density in the presence of a magnetic field $B= 2 \times 10^{19}~$Gauss to be, respectively, $M^{\rm max}\approx 1.14 M_\odot$, $R^{\rm max}\approx 4.75~{\rm km}$, and $\epsilon_c^{\rm max} \approx 41 \epsilon_{\rm sat}$. These values should be taken as extreme, since they correspond to assuming magnetic fields that yield an equation of state that sits in the causality limit. For comparison, the value ${\cal B}^{1/4}=145~$MeV for the bag constant produces quark stars with maximum mass $M_{\rm MIT}^{\rm max}\approx 2.01 M_\odot$ and radius $R_{\rm MIT}^{\rm max}\approx 10.9 ~{\rm km}$ in the absence of a magnetic field \cite{Schaffner-Bielich:2020psc}. Very large magnetic fields produce stars that are less massive, smaller and much more compact, in line with the findings of Ref. \cite{Ferreira:2022fjo}.

\section{Analytic pQCD model for the equation of state and quark magnetars}
\label{sec:pQCD-model}

\subsection{Analytic pQCD approximate equation of state}

\begin{figure*}[!ht]
\begin{subfigure}
 \centering
 \includegraphics[width=0.45\textwidth]{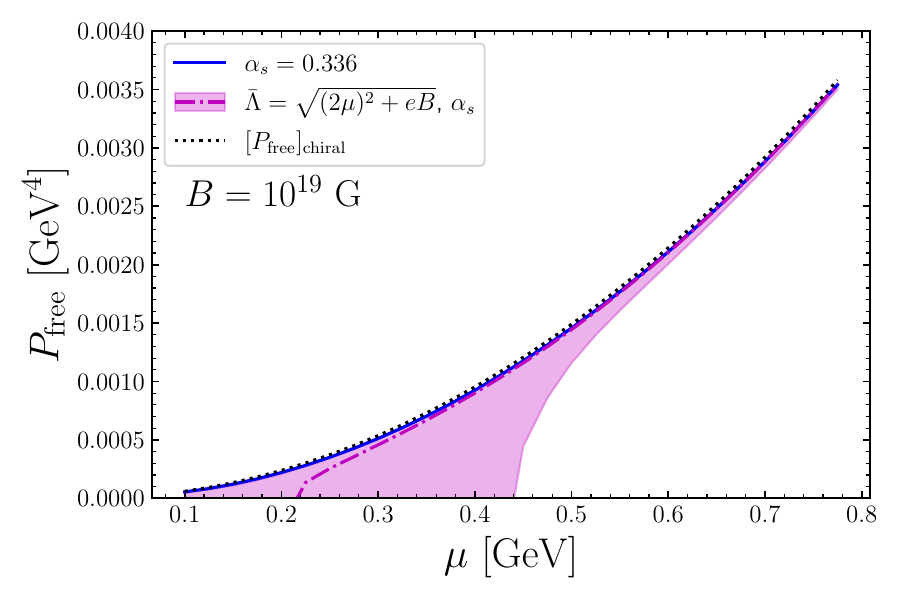} 
 \end{subfigure}
 \begin{subfigure}
 \centering
 \includegraphics[width=0.45\textwidth]{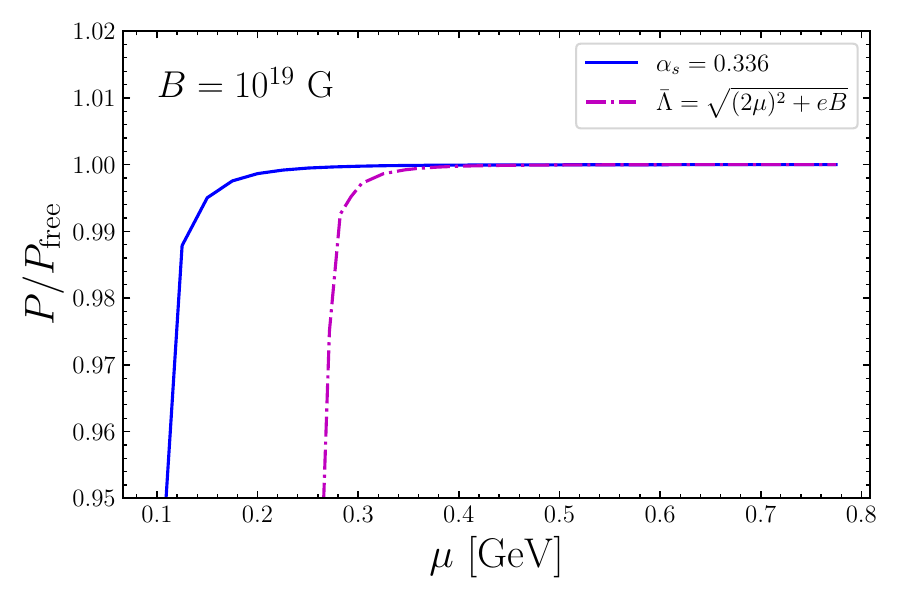} 
 \end{subfigure}
\caption{$P_{\rm free}$ (left) and $P/P_{\rm free}$ (right) as functions of the quark chemical potential for $B=10^{19}$ Gauss ($eB=0.059~$GeV$^2$). The bands correspond to changes in the central scale by a factor of $2$. For comparison, we also display the pressure in the chiral limit, given by Eq. (\ref{free-chiral}).}
\label{fig:Ptot_B00195}
\end{figure*}

As discussed previously, the contribution of the two-loop exchange diagram is negligible when compared to the one-loop free term in the case of very high magnetic fields. For magnetic fields expected to be found in the core of magnetars and higher, one can show that the exchange contribution remains small, but renormalization group running effects are relevant. This point is illustrated in Figure \ref{fig:Ptot_B00195}, where we show $P_{\rm free}$ and how it compares to the total pressure $P$ as functions of the chemical potential for $B=10^{19}$ Gauss, including the running bands that correspond to changes in the central scale by a factor of $2$. We also plot the pressure in the chiral limit, given by Eq. (\ref{free-chiral}), for comparison.

Hence, we can build a simple analytic pQCD-based model for the equation of state by using
\begin{equation}
 \frac{P_{\rm eff}}{N_c}=
 -\sum_f\frac{(q_fB)^2}{2\pi^2}\left[x_f\ln\sqrt{x_f}\right]+\sum_{f}\frac{(q_fB)}{4\pi^2}\left[ \mu_f P_F - m_f^2 \log\left( \frac{\mu_f+P_F}{m_f} \right) \right] \,,
  \label{eos-eff}
\end{equation}
including the renormalization group running. This represents an excellent approximation to the two-loop pQCD result at very large magnetic fields.

One will notice that we have larger bands for the pressure as we push pQCD towards smaller values of the magnetic field, essentially leaving the strict region of validity for the full calculation. The bands here provide a measure of the uncertainty in the perturbative calculation, a useful feature that is usually not present in model calculations. In fact, one can use the perturbative band at a given chemical potential to restrict possible equations of state, in the same fashion as performed in Ref. \cite{Kurkela:2014vha} for neutron star matter\footnote{Even though the exchange contribution is small for $B=10^{19}$ Gauss, such field is not high enough to justify the LLL approximation, as discussed previously. Hence, the equation of state above should not be naively applied to magnetars. Reliable results are only obtained for higher values of the magnetic field.}.

\subsection{Limits on quark magnetars from pQCD}

We can illustrate the applicability of the analytic approximate equation of state for cold quark matter in the presence of a high magnetic background obtained from perturbative QCD by computing the mass-radius diagram of quark magnetars assuming that the magnetic field is high enough. By doing that, we can place constraints on the behavior of the maximum mass and associated radius from perturbative QCD. Any given model description should approach these constraints for high enough values of $B$.

\begin{figure*}[!ht]
\begin{subfigure}
 \centering
 \includegraphics[width=0.45\textwidth]{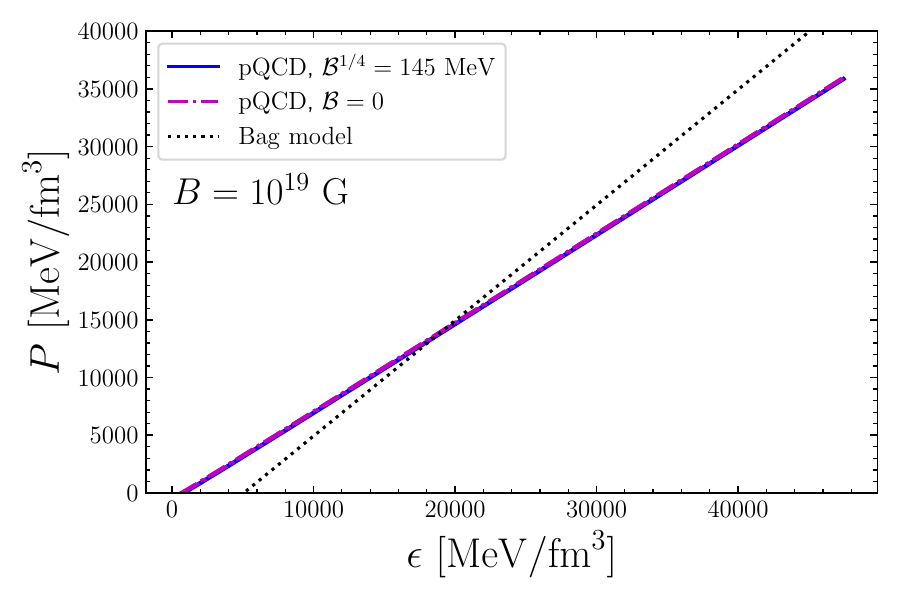} 
 \end{subfigure}
 \begin{subfigure}
 \centering
 \includegraphics[width=0.45\textwidth]{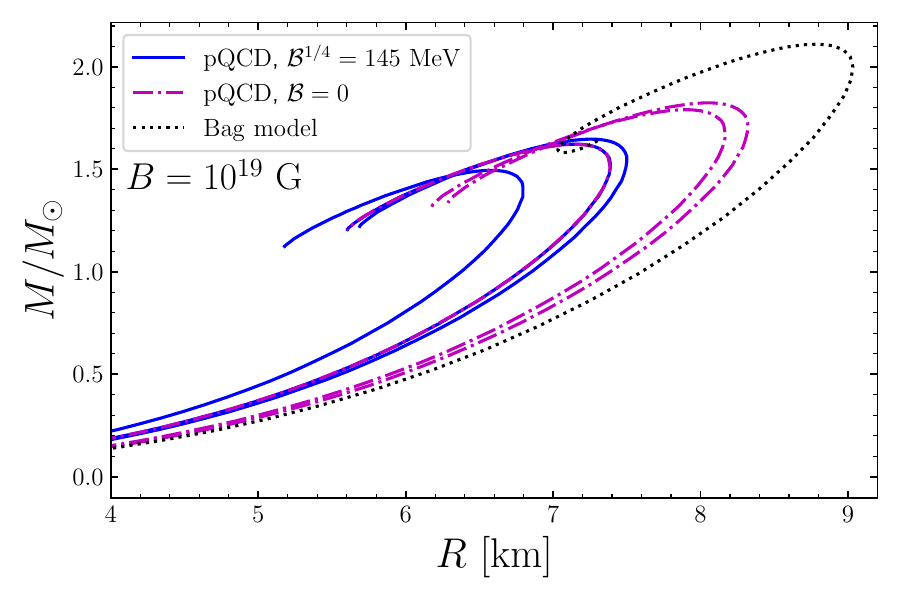} 
 \end{subfigure}
\caption{Equation of state (left) and mass-radius relation (right) obtained from perturbative QCD for $B=10^{19}$ Gauss.
We also show results from the magnetic bag model for comparison. For the perturbative equation of state, we exhibit results with and without a bag constant ${\cal B}^{1/4}=145~$MeV. The different curves correspond to changes in the central scale by a factor of $2$.}
\label{fig:Mvsr_B_10_19_G}
\end{figure*}

In order to describe quark magnetars, one must still impose beta equilibrium and charge neutrality. These are given, respectively, by the following conditions \cite{Schaffner-Bielich:2020psc}:
\begin{align}
    \mu_d=\mu_s=\mu_u+\mu_e
\end{align}
and
\begin{align}
    \frac{2}{3}n_u-\frac{1}{3}n_d-\frac{1}{3}n_s-n_e =0 \,.
\end{align}
Here $\mu_e$ is the electron chemical potential, 
\begin{align}
n_e=\frac{eB\mu_e}{2\pi^2}
\end{align}
is the electron number density, and 
\begin{align}
    n_f= N_c \frac{q_f B P_F}{2\pi^2} \label{density}
\end{align}
is the $f$-quark number density ($\mu_f > m_f$). The energy density of a magnetized system at zero temperature is given by
\begin{align}
    \epsilon=-P+\sum_f\mu_f \frac{\partial P_f}{\partial \mu_f} +\mu_e n_e  
    =&-P+\sum_f\mu_f n_f +\mu_e n_e \,.
\end{align}
%


Here, we should write the running of the strange quark mass in terms of the baryon chemical potential as $\bar{\Lambda}=\sqrt{(2\mu_B/3)^2+eB}$, where $\mu_B=\mu_u+\mu_d+\mu_s$. The fact that the scale is a function of the external parameter $\mu_B$ spoils the thermodynamic consistency. So, the pressure must be corrected to
\begin{align}
P=P_{\rm eff}+W-\frac{B^2}{2}+P_e \,,
\end{align}
where we add the purely magnetic contribution to the pressure, the free electron pressure,
\begin{align}
    P_e=\frac{eB\mu_e^2}{4\pi^2},
\end{align}
and a function ($W$) that ensures thermodynamic consistency, so that $(\partial P/\partial \mu_f)$ is equal to Eq. (\ref{density}) (see Refs. \cite{Gorenstein:1995vm,Wang:2000dc,Yin:2008me,Lenzi:2010mz,Restrepo:2022wqn,Ma:2023stj} for details). In our case, the $\mu_B$ dependence of the scale enters only in the strange pressure via the running of the mass. Thus, the function $W$ is given by
\begin{align}
\frac{\partial W}{\partial \mu_s}=-\frac{\partial (P_{\rm eff}+P_e)}{\partial m_s}\frac{\partial m_s}{\partial \bar\Lambda}\frac{\partial \bar\Lambda}{\partial \mu_s} \,.
%
\end{align}
%


In Figure \ref{fig:Mvsr_B_10_19_G} we show the equation of state, $P=P(\epsilon)$, and the mass-radius relation obtained from our simple analytic pQCD-based model for a magnetic field $B=10^{19}$ Gauss, an order of magnitude larger than a typical value in the interior of magnetars. 
We also show results from the magnetic bag model for comparison. Defining the dimensionless scale $X\equiv \bar{\Lambda}/\sqrt{(2\mu_B/3)^2+eB}$, the curves coming from perturbative QCD correspond to $X=1/2, 1, 2$, i.e., changes in the central scale by a factor of 2. We exhibit results with and without a bag constant (${\cal B}^{1/4}=145~$MeV). One can notice that the slope in the analytic pQCD-based model for the equation of state is smaller than the one from the magnetic bag model, so that the speed of sound is less than $1$ in the former. This is an effect of strong interactions via the renormalization group running and brings significant deviations from the magnetic bag model (cf. Figure \ref{fig:Ptot_B00195}) that affect the mass-radius relation lowering the values of the maximum mass and radius.

As expected from the analysis in the previous section, the masses and radii obtained are small when in the presence of a high magnetic field. For comparison, perturbative QCD to two-loops, and including the running of $\alpha_s$ and $m_s$, yields quark stars with maximum mass $M_{\rm pQCD}^{\rm max}\approx 2.16 M_\odot$ and radius $R_{\rm pQCD}^{\rm max}\approx 12 ~{\rm km}$ in the absence of magnetic field \cite{Fraga:2004gz}.

\begin{figure*}[!ht]
\begin{subfigure}
 \centering
 \includegraphics[width=0.45\textwidth]{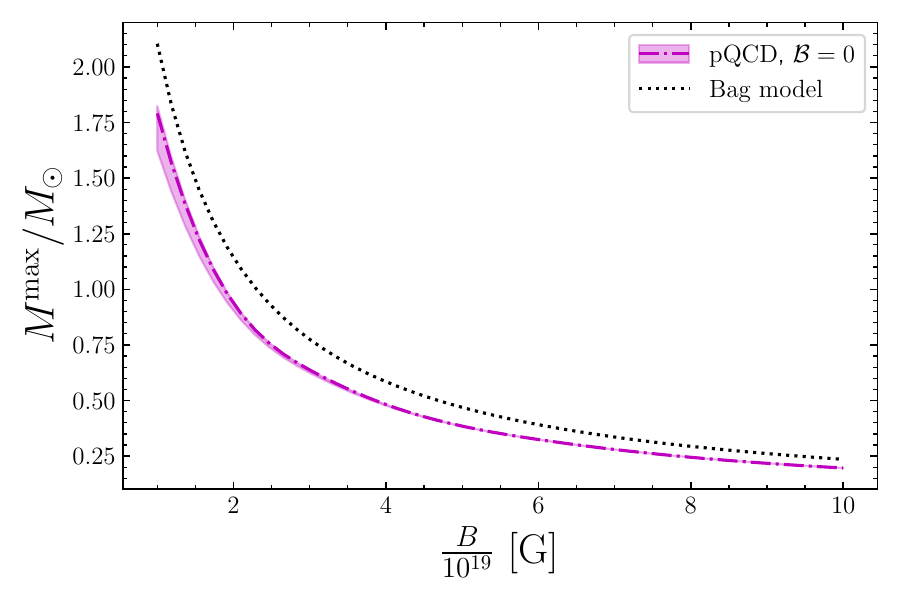} 
 \end{subfigure}
 \begin{subfigure}
 \centering
 \includegraphics[width=0.45\textwidth]{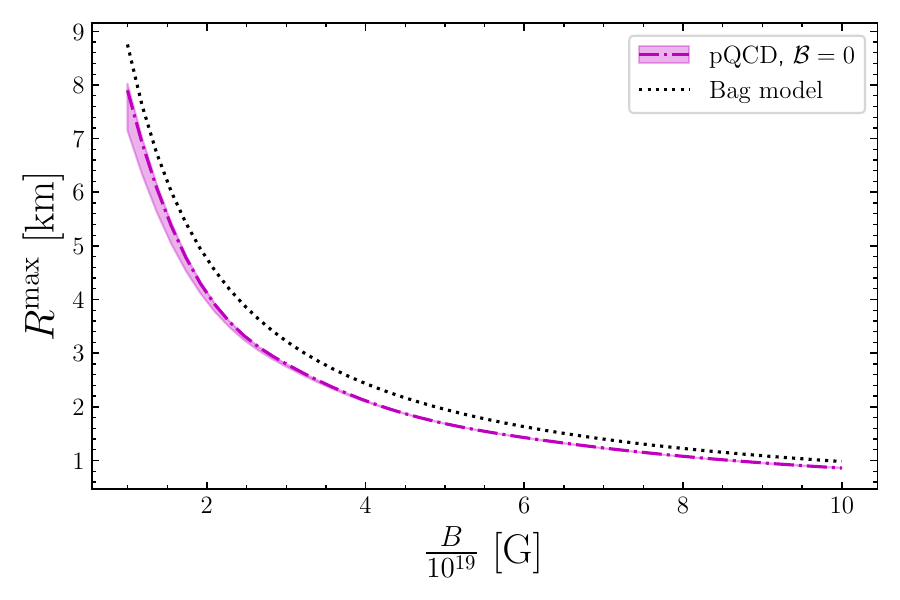} 
 \end{subfigure}
\caption{Maximum mass (left) and its respective radius (right) as functions of the magnetic field $B$ obtained from perturbative QCD. The different curves correspond to changes in the central scale by a factor of $2$. We also show results from the bag model for comparison.}
\label{fig:Mvsr_B0195}
\end{figure*}

Figure \ref{fig:Mvsr_B0195} displays the maximum mass and its respective radius as functions of the magnetic field $B$ obtained from perturbative QCD. The different curves correspond to changes in the central scale by a factor of $2$. In the figure, we also show results from the bag model for comparison. These results place constraints on the
behavior of the maximum mass and associated radius from perturbative QCD coming down from very high values of the magnetic field. Any given model description should, ideally, 
approach these constraints for high enough values of $B$.

Although the equation of state extracted from perturbative QCD cannot provide a good description of the low- and intermediate-density sectors all by itself, it is obtained from first principles and should guide the description of the high-density regime. For instance, model calculations of the equation of state for quark matter in the presence of strong magnetic fields should approach this behavior for high $\mu_B$ and very high $B$. To provide a more realistic description of the equation of state in magnetar matter, even for very large magnetic fields, one should match a low-density equation of state onto the equation of state from perturbative QCD, which is beyond the scope of this paper.

\section{Summary and outlook}
\label{sec:outlook}

In this paper we computed the pressure within perturbative QCD at finite density and very high magnetic fields up to two-loops and physical quark masses. Since we adopt the lowest-Landau level approximation in order to obtain analytic results and more control on qualitative aspects, the region of validity for our framework is restricted to $m_s \ll \mu_q \ll \sqrt{eB}$, where $m_s$ is the strange quark mass, $e$ is the fundamental electric charge, $\mu_q$ is the quark chemical potential, and $B$ is the magnetic field strength.

As also observed previously in the case of thermal magnetic QCD \cite{Blaizot:2012sd,Fraga:2023cef}, in the case of cold and dense QCD the contribution of the exchange diagram is essentially negligible. This sizable reduction in the exchange contribution in the presence of a very strong magnetic background might have remarkable effects on the convergence of the perturbative series. For instance, a clearly visible band related to the renormalization scale dependence appears only for relatively small values of the chemical potential.

We used this fact to build a simple (analytic) description for the high-density sector of the equation of state from our first-principle perturbative QCD result. It can be encoded in a simple formula which represents an excellent approximation to the two-loop pQCD result
at very large magnetic fields. One can use the associated perturbative band at a given chemical potential to restrict possible equations of state for magnetars, in the same fashion as performed in Ref. \cite{Kurkela:2014vha} for neutron star matter, provided that the magnetic field is high enough to justify the lowest-Landau level approximation.

If one considers the chiral limit in an extremely large magnetic background, a much stiffer effective magnetic bag model emerges, producing a Zel'dovich equation of state that sits in the causal limit. The reason is that one has $P_{\rm bag}\sim eB \mu^2$ in the magnetic bag model, instead of the usual $P_{\rm MIT}\sim \mu^4$ behavior. Within this model, we estimated the maximum mass, radius and central energy density for very large magnetic fields.

As an illustration, we used our simple analytic pQCD-based model to compute the equation of state $P=P(\epsilon)$ for magnetized quark star matter, imposing beta equilibrium and charge neutrality. Then, we obtained the mass-radius diagram for quark magnetars by solving the TOV equations. All our results have renormalization-group scale dependence, which provides at least a rough measure of their uncertainty. We also computed the maximum mass and its respective radius as functions of $B$, 
placing constraints on their behavior from perturbative QCD coming down from very high values of the magnetic field. Model calculations of the equation of state for quark matter in the presence of strong magnetic fields should approach this behavior for high densities and extreme values of $B$.

Of course, a more realistic picture of magnetars should incorporate the low-density sector and the matching of the equations of state, besides the inclusion of crust effects and a more thorough treatment of the magnetic field profile. This would allow for the computation of magnetic effects on other relevant observables, such as the tidal deformability. 

From the analysis of the perturbative series of magnetic QCD performed here for the case of cold quark matter, and its thermal counterpart in Refs. \cite{Blaizot:2012sd,Fraga:2023cef}, the convergence in powers of $\alpha_s$ seems to be much more well behaved than in the case with no magnetic field. The challenging technical extension that would permit more direct applications to astrophysical phenomenology would be going beyond the lowest-Landau level approximation. 


\begin{acknowledgments}
This work was partially supported by CAPES (Finance Code 001), Conselho Nacional de Desenvolvimento Cient\'{\i}fico e Tecnol\'{o}gico (CNPq), Funda\c c\~ao Carlos Chagas Filho de Amparo \` a Pesquisa do Estado do Rio de Janeiro (FAPERJ), and INCT-FNA (Process No. 464898/2014-5). T.E.R acknowledges support from FAPERJ, Process SEI-260003/019683/2022. This work has been supported by STRONG-2020 ``The strong interaction at the frontier of knowledge: fundamental research and applications'' which received funding from the European Union’s Horizon 2020 research and innovation program under grant agreement No 824093.
\end{acknowledgments}
\bibliography{refs.bib}
\end{document}